\documentclass[aps]{revtex4}  
\usepackage[latin1]{inputenc}
\usepackage{amsmath}
\usepackage{amsfonts}
\usepackage{amssymb}
\usepackage{graphicx,color}

\begin{document}

\title{Imaging Spatial Quantum Correlations through a Scattering Medium}

\author{Soro Gnatiessoro $^{1,*}$, Alexis Mosset$^1$, Eric Lantz$^1$, and Fabrice Devaux$^1$}

\address{$^1$ Institut FEMTO-ST, D\'epartement d'Optique P. M. Duffieux, UMR 6174 CNRS \\ Universit\'e Bourgogne Franche-Comt\'e, 15b Avenue des Montboucons, 25030 Besan\c{c}on - France}

\date{\today}
\email[Corresponding author:]{gnatiessoro.soro@femto-st.fr}



\begin{abstract}
 We image with cameras entangled photon light transmitted through a random medium. Near-field and far-field spatial quantum correlations show that entangled photon pairs (bi-photons) generated by spontaneous optical parametric down-conversion exhibit speckle pattern. In contrast, no information from the random medium can be extracted using incoherent light issued from one photon of the pair. Although these measurements require several hours to record thousands of image pairs, our method is instantaneous for the recording of one pair of twin images and involve all the photons of the images.
\end{abstract}


\maketitle

\section{Introduction}
Quantum correlations between optical fields produced in spontaneous parametric down conversion (SPDC) were first observed in 1970 by Burnham and Weinberg \cite{PhysRevLett.25.84,walborn_spatial_2010}. Since the first demonstration of correlation imaging and interference with entangled source \cite{PhysRevA.52.R3429}, spatial correlations have been a hot topic in the area of imaging \cite{devaux_spatial_2000,lantz_numerical_2001,brambilla_simultaneous_2004,jedrkiewicz_quantum_2006}. More recently, an experiment done in our group allowed the spatial information stored in a phase hologram to be retrieved \cite{devaux_quantum_2019} by measuring quantum spatial correlations between two images formed by spatial entangled twin photons. In the last years, more studies dealt with the propagation of entangled two photons states through random medium. This phenomenon has been both theoretically \cite{PhysRevLett.81.1829,skipetrov_quantum_2007,beenakker_two-photon_2009,cande_quantum_2013,cande_transmission_2014,li_statistical_2019} and experimentally \cite{smolka_observation_2009,peeters_observation_2010,defienne_adaptive_2018,defienne_spatially-entangled_2019} investigated. From the theoretical point of view, in 1998, Beenakker \cite{PhysRevLett.81.1829} introduced a convenient formalism of the input-output relations between the incoming and outgoing quantum states. Experimentally, in 2009, Smolka $et\,al$. \cite{smolka_observation_2009} reported the spatial observation between the scattered modes in the random medium and this study showed that the strength of the quantum correlation is related to the number of incident photons. Moreover, in 2010, Peeters $et\,al$. \cite{peeters_observation_2010} showed that, after propagation of entangled photon pairs through a random medium, the spatial repartition of the coincidence counts rate exhibits a speckle structure which cannot be obtained with the single count rate. More recently, Defienne $et\,al$. \cite{defienne_spatially-entangled_2019} reported the generation of entangled photon pairs with a tunable degree of spatial entanglement by controlling the spatial coherence of the pump beam with random diffusers. While in \cite{peeters_observation_2010}  two ponctual detectors are scanned over different positions, we propose in this paper to directly image the two detection planes using two electron-multiplying charge coupled device (EMCCD) cameras able to detect single photons \cite{lantz_multi-imaging_2008}. Obviously, since all photons are detected, correlations are evidenced on the whole set of photons. Here two configurations will be investigated. First, we put a thin diffuser in the crystal near-field and correlations are observed in the far-field of the crystal. Second, a thin diffuser is placed in the crystal far-field and near-field correlations are observed. Finally, we performed stochastic simulations \cite{lantz_spatial_2004} to confirm the experimental results and the spatial structure expected in the theoretical expression of the correlation function, both in the far-field and near-field correlations.

\section{Optical speckle pattern with entangled photon pairs }
To describe the propagation of light through  a random medium, we consider the two configurations illustrated in Fig. 1.
\begin{figure}[htbp]
	\centering
	\includegraphics[width=12cm]{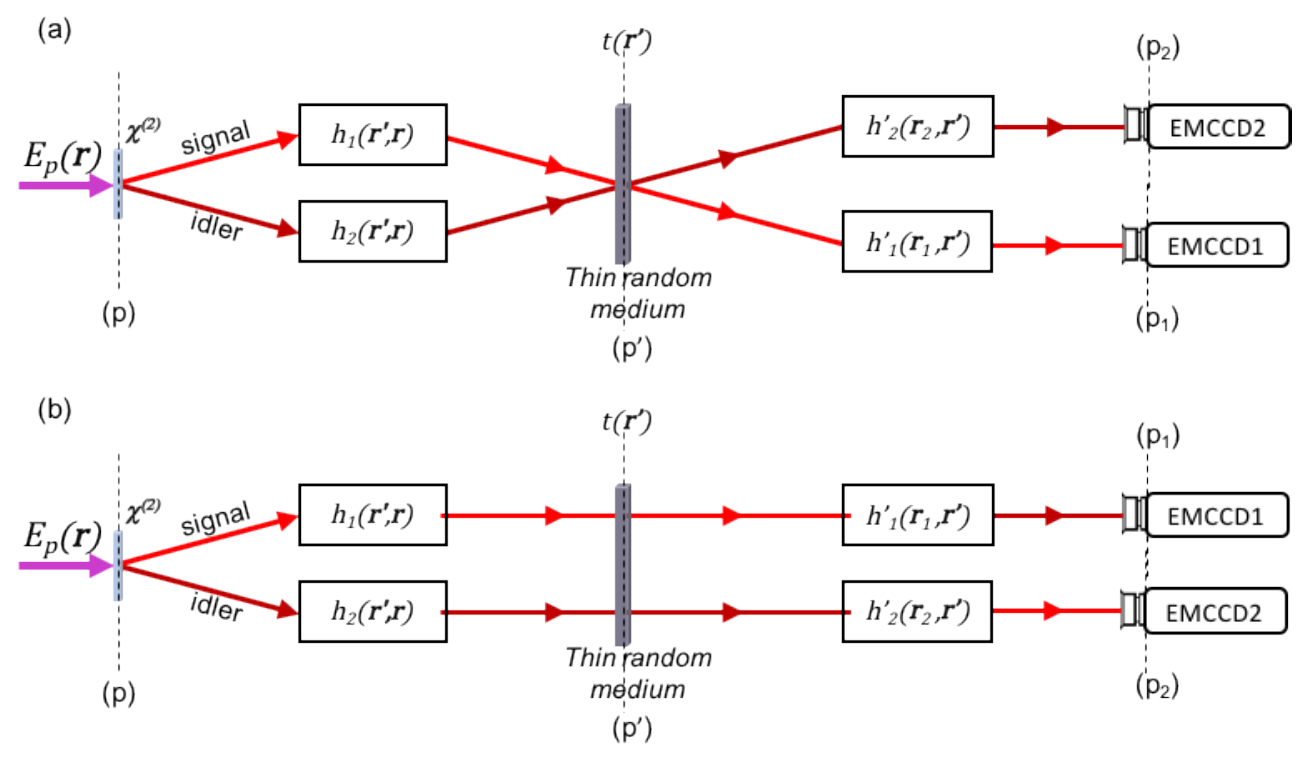}
	\caption{Two photon speckle generation with entangled photon pairs. The two configurations (a) and (b) are related to the far-field and near-field correlations, respectively. The $\chi ^{(2)}$ nonlinear crystal is pumped by laser pulses at the angular frequency $\omega_p$. The entangled photons (signal and idler) are generated at $\omega_s$ and $\omega_i=\omega_p-\omega_s$. Two separated optical systems with the transfer functions $h_1$ and $h_2$ are used to focus the signal and idler beams onto a thin random medium that transmits the two beams with the transmissions $t_s$ and $t_i$ for the signal and idler beams respectively.Then, another two separated optical systems with the transfer functions $h_{1}'$ and $h_{2}'$ are used to image the transmitted beams on the $EMCCD1$ and $EMCCD2$ cameras.}\label{fig1}
\end{figure}
The output face of the thin nonlinear crystal lies in the plane (P), with a transverse vector $\textbf{r}$. The thin random medium, that will be assumed as a pure phase-object of transmission  $t(\textbf{r'})=e^{i\Phi(\textbf{r'})}$ where $\Phi(\textbf{r'})$ is the phase modulation in the transverse plane. It lies in a plane$(P')$ with a transverse vector $\textbf{r'}$. The detectors arrays $EMCCD1$ and $EMCCD2$  lie in the planes $(P_1)$ and $(P_2)$ with transverse vectors $\textbf{r}_{1,2}$.
Let us consider a SPDC light emitted from a planar thin nonlinear crystal illuminated by a monochromatic beam of frequency $\omega_p$ and of amplitude $E_{p}(\textbf{r})$. For a sufficiently thin crystal, we assume that the two photons of the pair are created at the same random place in the nonlinear crystal \cite{devaux_quantum_2019}.
First, let us name $t_s(\textbf{r'})$ and $t_i(\textbf{r'})$ the transmission of the scattering medium for the signal and idler beams. Then, we can write the impulse-response functions of the imaging systems between the crystal and the detection planes for the signal and the idler as \cite{abouraddy_entangled-photon_2002}:

\begin{eqnarray} \label{eq1}
\left\{\begin{array}{cl}
h_{s}\left(\textbf{r}_{1},\textbf{r},\omega_{s}\right)=\int d\textbf{r}_{s}'h_{1}\left(\textbf{r}'_{s},\textbf{r},\omega_{s}\right)t_{s}\left(\textbf{r}'_{s},\omega_{s}\right)h'_{1}\left(\textbf{r}_{1},\textbf{r}'_{s},\omega_{s}\right)\\
h_{i}\left(\textbf{r}_{2},\textbf{r},\omega_{p}-\omega_{s}\right)=\int d\textbf{r}'_{i}h_{2}\left(\textbf{r}'_{i},\textbf{r},\omega_{p}-\omega_{s}\right)t_{i}\left(\textbf{r}'_{i},\omega_{p}-\omega_{s}\right)h'_{2}\left(\textbf{r}_{2},\textbf{r}'_{i},\omega_{p}-\omega_{s}\right)\end{array}\right.
\end{eqnarray}
Where $h_{1,2}(\textbf{r}',\textbf{r})$ are the impulse-response functions describing the propagation from the crystal plane up to the plane of the random medium. Similarly, $h'_{1}(\textbf{r}',\textbf{r}_{1})$ and $h'_{2}(\textbf{r}',\textbf{r}_{2})$ describe the propagation from the random medium to the cameras. If  we assume that SPDC is detected at the degeneracy $\left(\omega_{i}=\omega_{s}=\omega_{p}/2\right)$ with a narrow band interferential filter, we can neglect chromatic dispersion of the whole optical systems. Then, the joint probability of detection of photons at both detectors arrays $EMCCD1$ and $EMCCD2$ is given by \cite{saleh_duality_2000}:

\begin{equation} \label{eq2}
\psi(\textbf{r}_{1},\textbf{r}_{2})\propto \int E_{p}(\textbf{r})h_{s}(\textbf{r}_{1},\textbf{r})h_{i}(\textbf{r}_{2},\textbf{r})d\textbf{r}
\end{equation}

Two imaging configurations are considered in this paper. In the first one (Fig. 1a), the scattering medium lies in the near-field of the crystal and the detectors arrays, in the far-field, measure momentum correlations. In the second imaging configuration(Fig. 1b), the scattering medium lies in the far-field of the crystal and detectors arrays image the near-field. In this later case, position correlations are measured.

\subsection{Scattering medium in the near-field, far-field correlations}
In this imaging configuration where momentum correlations are measured, a $4-f$ optical system is used in both signal and idler branches to image the signal and idler beams at the same place of the random medium. Assuming that the phase matching angular bandwidth allows all the photons from the signal and idler beams to be collected by the $4-f$ optical system, we can neglect the effect of the $4-f$ system. In this case, we can consider that the transmission for both beams verifies: $t_s(\textbf{r}'_{s})=t_{i}(\textbf{r}'_{i})=t(\textbf{r}')$. Then, two identical lenses are used to record the signal and idler beams onto the detectors arrays $EMCCD1$ and $EMCCD2$. Thereby, $h_{1}'$ and $h_{2}'$ correspond to a two identical $2-f$ impulse-response functions which take the form \cite{simon_quantum_2016}:

\begin{eqnarray} \label{eq3}
\left\{\begin{array}{cl}
h_{1}'(\textbf{r}_{1},\textbf{r}')=\frac{e^{-2ikf}}{i\lambda f}e^{\frac{-ik}{f}\textbf{r}_{1}.\textbf{r}'}\\
h_{2}'(\textbf{r}_{2},\textbf{r}')=\frac{e^{-2ikf}}{i\lambda f}e^{\frac{-ik}{f}\textbf{r}_{2}.\textbf{r}'}\end{array}\right.
\end{eqnarray}

where $k$ and $\lambda$ are the signal/idler wave number and wavelength at the degeneracy and $f$ the focal length of the two lenses.
Finally, the impulse-response functions can be simply expressed in the form:

\begin{eqnarray} \label{eq4}
\left\{\begin{array}{cl}
h_{s}(\textbf{r}_{1},\textbf{r}')=t(\textbf{r}')\frac{e^{-2ikf}}{i\lambda f}e^{\frac{-ik}{f}\textbf{r}_{1}.\textbf{r}'}\\
h_{i}(\textbf{r}_{2},\textbf{r}')=t(\textbf{r}')\frac{e^{-2ikf}}{i\lambda f}e^{\frac{-ik}{f}\textbf{r}_{2}.\textbf{r}'}\end{array} \right.
\end{eqnarray}

In this case, the detection joint probability becomes:

\begin{equation} \label{eq5}
\psi(\textbf{r}_{1},\textbf{r}_{2})\propto\int E_{p}(\textbf{r}')t^{2}(\textbf{r}')e^{\frac{-ik}{f}(\textbf{r}_{1}+\textbf{r}_{2}).\textbf{r}'}d\textbf{r}'
\end{equation}

The correlation function corresponding to the detection of the two photons (signal and idler) on the $EMCCD1$ and $EMCCD2$ cameras is given by \cite{saleh_duality_2000,devaux_quantum_2019}:

\begin{equation} \label{eq6}
G^{(2)}(\textbf{r}_{1},\textbf{r}_{2})=\left|\psi\left( \textbf{r}_{1},\textbf{r}_{2}\right)\right|^{2}\propto\left| \tilde{E}_{p}\left(2\pi\frac{\textbf{r}_{1}+\textbf{r}_{2}}{\lambda f}\right)\ast\widetilde{t^2}\left(2\pi\frac{\textbf{r}_{1}+\textbf{r}_{2}}{\lambda f}\right)\right|^{2}
\end{equation}

Where $*$ denotes the convolution product and the tilde the bidimensional Fourier-transform operator. Eq.(6) shows that the correlation function corresponds to a convolution product between the pump spatial spectrum and the speckle issued from the random medium. Therefore, we expect a speckle pattern in the correlation image in this case inasmuch as the pump beam is sufficiently large in the direct space. 

\subsection{Scattering medium in the far-field, near-field correlations}
In this configuration, the signal and idler beams are transmitted through two sequential and identical $2-f$ optical systems with a thin random medium in the intermediate Fourier plane between them. The random medium includes two non-identical transmission $t_{s}(\textbf{r}'_{s})$ and $t_{i}(\textbf{r}'_{i})$ since the far-field of the signal and idler beams is transmitted by two different areas of the random medium ($\textbf{r}'_{s}\neq\textbf{r}'_{i}$). Therefore, we can consider that the signal and idler branches comprise a $4-f$ optical with apertures $t_{s}(\textbf{r}'_{s})$ and $t_{i}(\textbf{r}'_{i})$, respectively. Under these considerations, the impulse-response functions in the two branches can be expressed as \cite{saleh_duality_2000,simon_quantum_2016}:

\begin{eqnarray} \label{eq7}
\left\{\begin{array}{cl}
h_{s}(\textbf{r}_{1},\textbf{r})=\frac{e^{-4ikf}}{i\lambda f}\tilde{t}_{s}\left( 2\pi\frac{\textbf{r}-\textbf{r}_{1}}{\lambda f}\right)\\
h_{i}(\textbf{r}_{2},\textbf{r})=\frac{e^{-4ikf}}{i\lambda f}\tilde{t}_{i}\left( 2\pi\frac{\textbf{r}-\textbf{r}_{2}}{\lambda f}\right)\end{array}\right.
\end{eqnarray}

Where $\tilde{t}_{s}$ and $\tilde{t}_{i}$ are the Fourier-transform of $t_{s}(\textbf{r}')$ and $t_{i}(\textbf{r}')$ respectively. From Eq. (3), the joint probability can be expressed as:

\begin{equation} \label{eq8}
\psi(\textbf{r}_{1},\textbf{r}_{2})\propto\int E_{p}(\textbf{r})\tilde{t}_{s}\left( 2\pi\frac{\textbf{r}-\textbf{r}_{1}}{\lambda f}\right)\tilde{t}_{i}\left( 2\pi\frac{\textbf{r}-\textbf{r}_{2}}{\lambda f}\right)d\textbf{r}
\end{equation}
In the case, the correlation function becomes:

\begin{equation} \label{eq9}
G^{(2)}(\textbf{r}_{1},\textbf{r}_{2})=\left| \psi(\textbf{r}_{1},\textbf{r}_{2})\right|^{2}\propto\left|\int   E_{p}(\textbf{r})\tilde{t}_{s}\left( 2\pi\frac{\textbf{r}-\textbf{r}_{1}}{\lambda f}\right)\tilde{t}_{i}\left( 2\pi\frac{\textbf{r}-\textbf{r}_{2}}{\lambda f}\right) d\textbf{r}\right|^{2}
\end{equation}
This equation corresponds to the magnitude of a triple correlation function between the near-field pump amplitude and the Fourier transform of two random uncorrelated phase functions \cite{abouraddy_entangled-photon_2002}. Due to the complexity of this equation, it is not simple to predict the spatial structure expected in the correlation function. In this case, we performed stochastic simulations to confirm the spatial structure obtained in our experimental results.

\section{experimental setup and results}
\subsection{Far-field correlation: experimental setup}
The experimental setup in this subsection is identical to that described in \cite{devaux_quantum_2019}, where the hologram is replaced by a thin diffuser.

\begin{figure}[htbp]
	\centering
	\includegraphics[width=12cm]{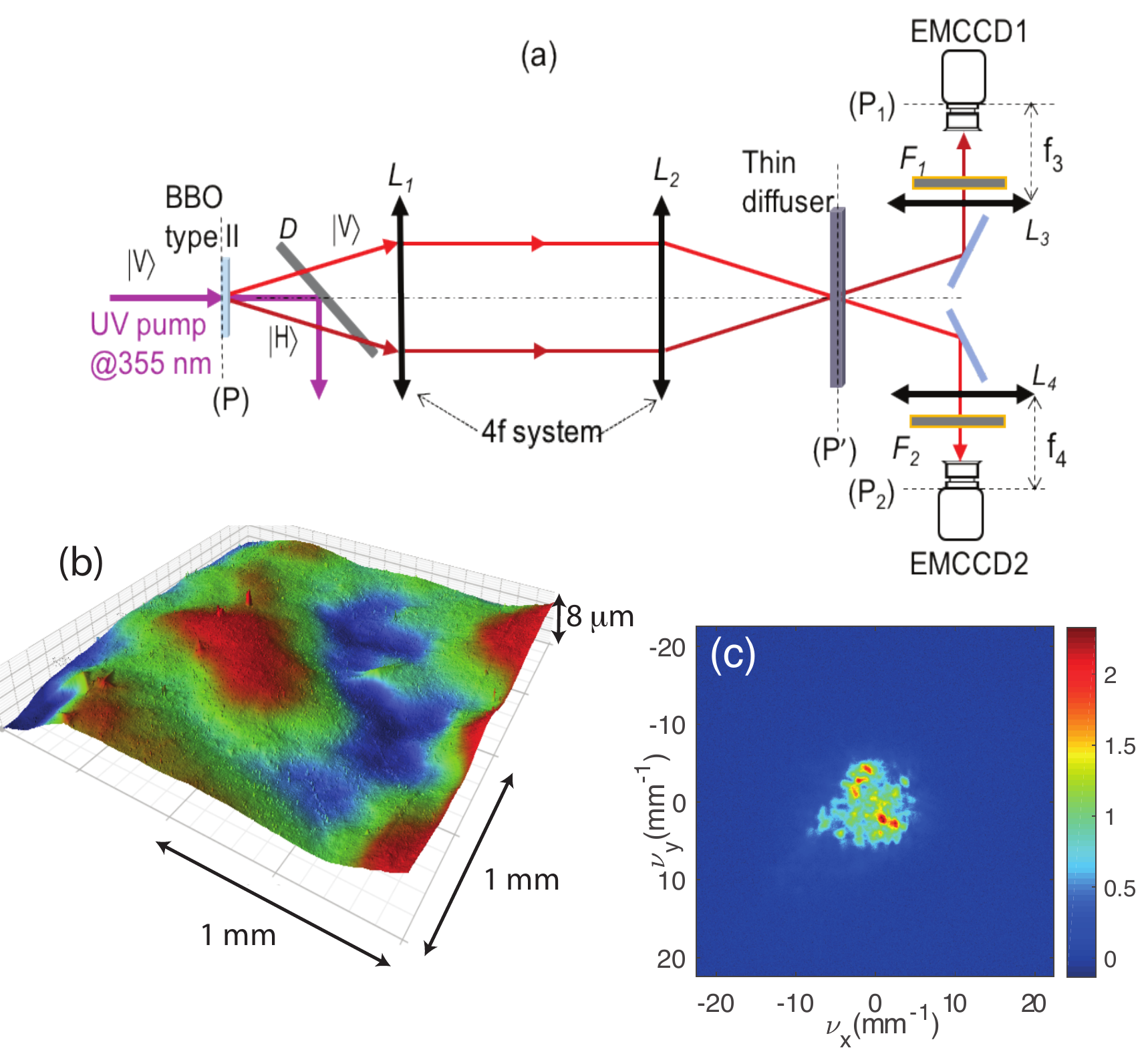}
	\caption{(a) Experimental setup: Entangled photon pairs at 710 $nm$  are generated via SPDC in a type-II BBO using a 330 $ps$ pump pulse at 355 $nm$. The crystal is imaged onto the thin diffuser with a $4-f$ optical system. The entangled photon pairs (signal and idler) transmitted by the diffuser are detected and resolved spatially in the far-field on two EMCCD cameras. (b): 3D-profile of the diffuser and (c): Image of the laser scattered light in the far-field. $ \rvert V\rangle$ and $\rvert H\rangle$ represent the vertical and horizontal polarizations. $(P_1)$ and $(P_2)$ represent the Fourier plane and $(P')$ the image plane. D is a dichroic mirror, $F_3$ and $F_4$ are the interferential filters. }\label{fig2}
\end{figure}

{Fig. 2a shows the experimental setup. Entangled photon pairs are generated by type-II SPDC in a 0.8 $mm$ long $\beta$-barium borate (BBO) crystal pumped by a collimated pulsed laser at 355 $nm$. The pump pulses (330 ps) are provided by a passively Q-switched Nd:YAG laser (27 $mW$ mean power, $1kHz$ repetition rate and 1.6 $mm$ FWHM beam diameter) see \cite{devaux_quantum_2019}. The lenses $L_1$ ($f_1$=75 $mm$) and $L_2$ ($f_2$=75 $mm$) ($4-f$) imaging system are used to image the output face of the crystal onto the diffuser. The entangled photon pairs transmitted by the diffuser are separated thanks to walk-off. Far-field detection is performed by two EMCCD cameras lying in the focal plane of the lenses $L_3$ ($f_3$=150 $mm$) and $L_4$ ($f_4$=150 $mm$). Before detection, the SPDC beams are filtered around degeneracy with narrow-band interferential filters $F_1$ and $F_2$ ($@710\,nm, \Delta\lambda=4\,nm$). Fig. 2b shows a 3D-profile of an area of $2\times2\,mm^2$ of the diffusers. The scattering medium is made of a microscopic slide with one side attacked with fluorydic acid. This process allows the production of diffusers of deep roughness ($\sim 3\,\mu m$) and large waviness profiles (few 100 $\mu m$). The deep roughness profile ensures a spatial phase modulation of the transmitted beams with a large amplitude $(>4\pi)$ and the large waviness profile ensures that most of the scattered light is spread over a small angular aperture, smaller than the phase matching angular bandwidth (47 $mrad$). Consequently, we assume that most of the scattered twin photons are collected by the imaging system.  To ensure that the diffuser is in the image plane of the crystal, we used a He-Ne laser (no SPDC in this regime) to check the alignment of all optical components and to minimize defocusing of the diffuser plane. Fig. 2c shows the experimental image recorded in the far-field of the diffuser using the He-Ne laser which provides a coherent beam. In Fig. 2c, the speckle pattern carries information both on the coherence properties of the laser light and microscopic details of the diffuser \cite{peeters_observation_2010}. We determined the size of the speckle pattern and the size of the scatterers along transverse direction. Fig. 2c shows the speckle pattern obtained with a FWHM of 8 $mm^{-1}$ in the far-field that corresponds to an average size of the diffuser waviness of 125 $\mu\,m$, in agreement with the waviness of the surface of the diffuser (see Fig. 2b)}.
Now, we discuss the case where SPDC is used.

\begin{figure}[htbp]
	\centering
	\includegraphics[width=13cm]{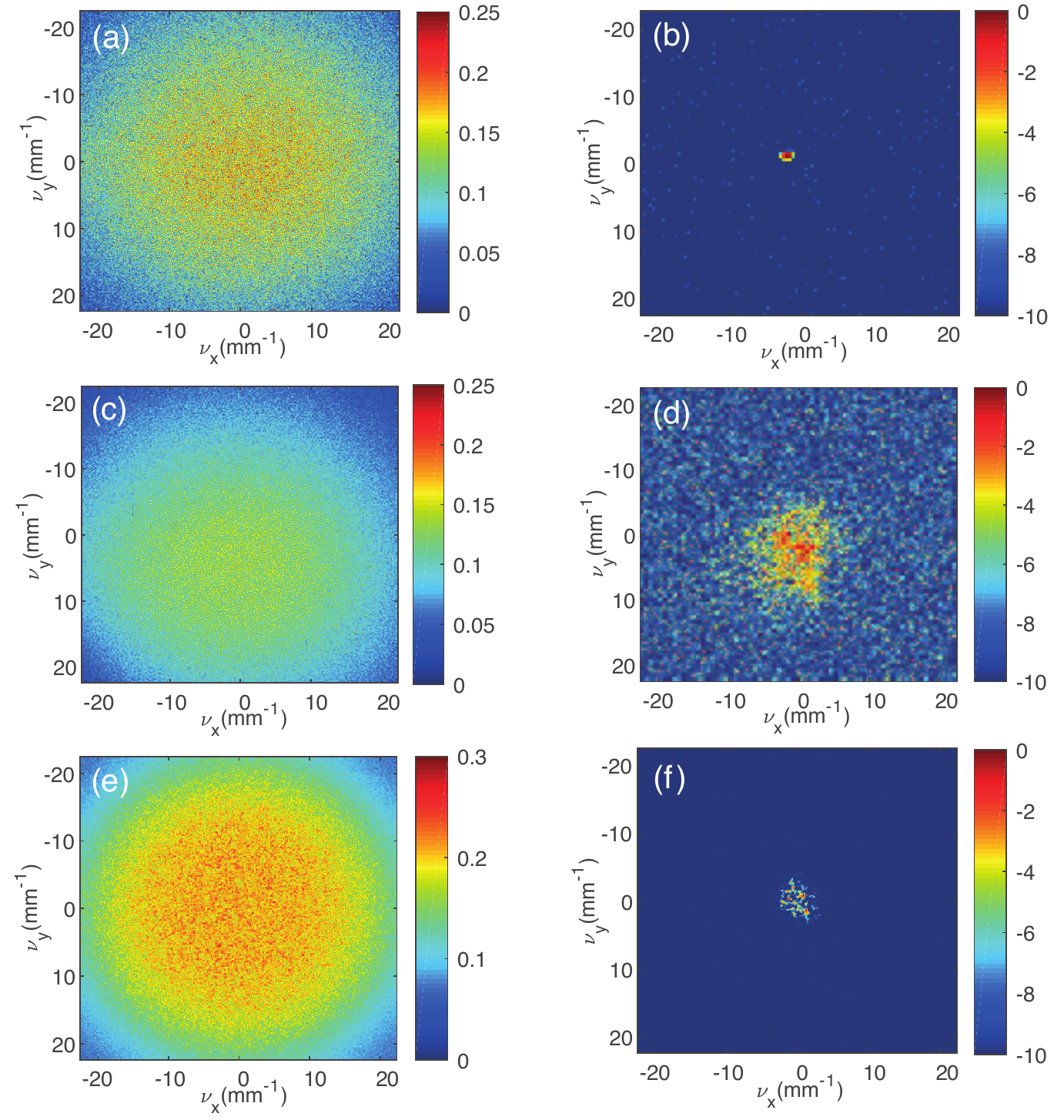}
	\caption{Without diffuser,(a) average photon number in single far-field images (signal or idler) of SPDC and (b) measured correlation function in dB between 100 twin images. With diffuser, (c) average photon number in single far-field images (signal or idler) of SPDC and (d) measured correlation function in dB over 40 000 twin images. With stochastic simulations, (e) average photon number in single far-field images (signal or idler) of SPDC and (f) correlation function issued from 10 000 stochastic simulations with diffuser.}\label{fig3}
\end{figure}

First, we measure the correlation function with a set of 100 twin images without the diffuser in order to characterize the bi-photon state. Fig. 3a shows a typical single far-field image of SPDC simply conditioned by phase matching. As shown in \cite{peeters_observation_2010}, the absence of a speckle pattern in the single images of SPDC results from the incoherent character of the light formed by a single beam of the entangled light. The momentum spatial correlation function is calculated by summing the cross-correlations of the signal image of a pair with the 180° rotated idler image of the same pair. Fig. 3b shows a narrow correlation peak in the correlation function without the diffusing medium. This correlation peak gives access to the degree of correlation, i.e the ratio between  the number of photons  detected in pairs  and the total number of photons \cite{devaux_quantum_2019}. In this experiment, the value of the degree of correlation without diffuser is equal to 0.22. This value is close to the effective quantum efficiency $0.26$ of the entire detection system and the integral of the correlation peak $0.23$ obtained in \cite{lantz_einstein-podolsky-rosen_2015}. The slight difference can be attributed to the additional losses induced by the two supplementary lenses necessary to image the diffusing medium. By approximating the correlation peak  to a two dimensional Gaussian profile, we can calculate the conditional  variances which correspond to the widths of the normalized cross correlations peak along the $x$ and $y$ axis. For more details about the calculation of the conditional variances, see \cite{phdthesis}. From the normalized cross correlation in momentum calculated with a set of 100 twin images (see Fig. 3b), we calculated the conditional variances and we obtained $\sigma_{\nu_x}^2=5,00.10^{-5}\,\mu m^{-2}\hbar^2$ along $x$ axis and $\sigma_{\nu_y}^2=1,05.10^{-5}\,\mu m^{-2}\hbar^2$ along $y$ axis.
Secondly, we consider the diffuser in the image plane of the crystal as shown in the experimental setup (see Fig. 2a). Now we consider a single far-field image (signal or idler) of SPDC. Similarly to fig. 3a without diffuser, Fig. 3c and Fig. 3e show typical single far-field images of SPDC simply conditioned by phase matching. Moreover, Fig. 3a; Fig. 3c and Fig. 3e show that the profile of the SPDC beam scattered by the diffuser is not significantly widened. For the correlation function, Fig. 3d and Fig. 3f show a speckle pattern. The presence of the speckle pattern is in agreement with the predictions of Eq. (6) and also consistent with the works done by Peeters $et\,al.$ \cite{peeters_observation_2010} where, by scanning the positions of both detectors independently in the horizontal line, a far-field correlation peak appears around a null value of the sum coordinates. As we did with the laser beam in fig. 2c, we determined the speckle pattern size in Fig. 3d and we get a FWHM of 8,4 $mm^{-1}$ that corresponds to an average size of the diffuser waviness of 119 $\mu m$, in agreement with the waviness of the surface the diffuser. The degree of correlation with diffuser estimated in Fig. 3d is 0.17. To explain the difference between the degree of correlation without diffuser and with diffuser, we have considered that a part of correlations are lost due to the reflection or absorption effects induced by the diffuser and also to the fact that few scattered photons are not collected by the imaging system. 

\subsection{Near-field correlation: experimental setup}
In this second part, we report the near-field correlations between entangled photon of pairs transmitted through the diffuser. As mentioned in section II.B, the far-field of the crystal is imaged onto the diffuser and the spatial correlations are measured between near-field images of the crystal. The experimental setup is depicted in Fig. 4.

\begin{figure}[htbp]
	\centering
	\includegraphics[width=12cm]{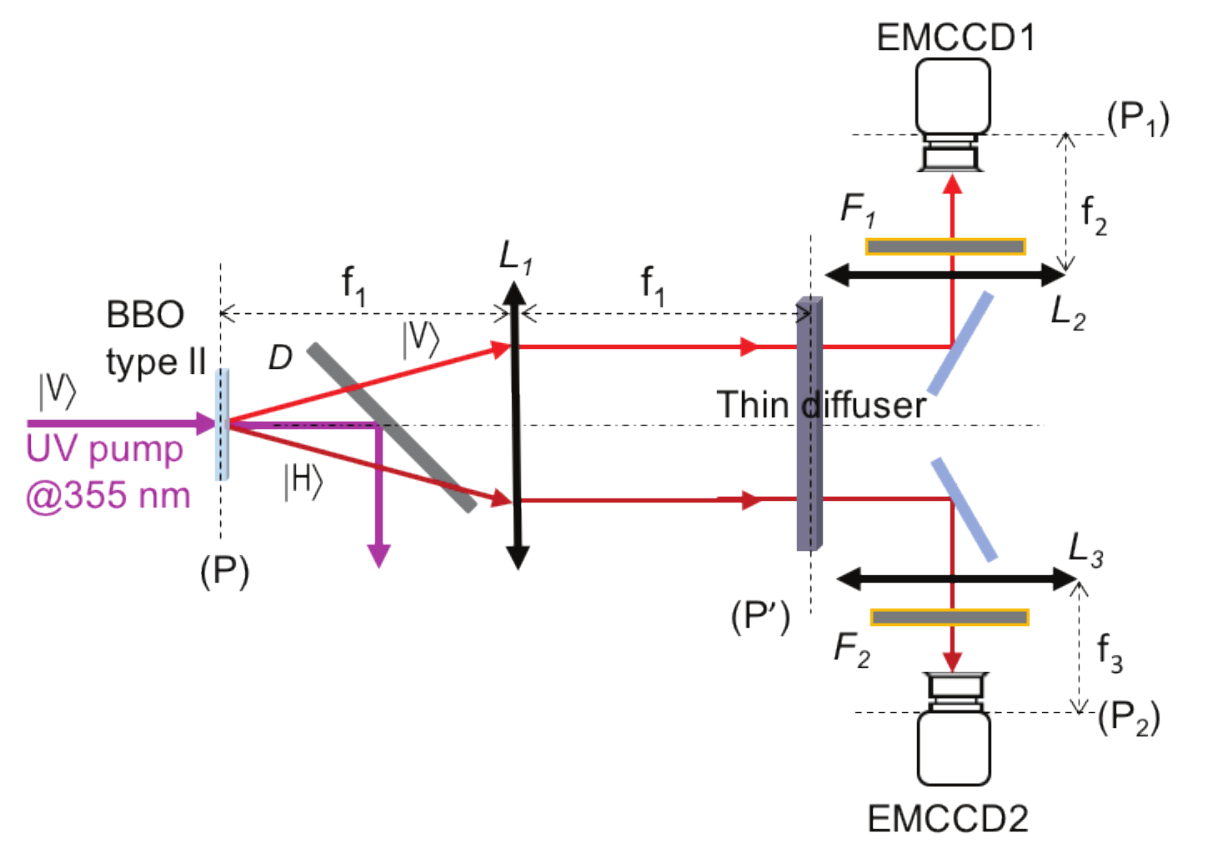}
	\caption{Experimental setup:Entangled photon pairs at 710 $nm$ are generated via SPDC in a type-II BBO using 330 $ps$ pump pulse at 355 $nm$. Lens $L_1$ ($f_1$=150 $mm$) is used to image the far field of the crystal onto the diffuser placed in the Fourier plane. The entangled photon pairs (signal and idler) transmitted by the diffuser are detected in the near field on the EMCCD1 and EMCCD2 cameras using lenses $L_2$ ($f_2$=150 $mm$) and $L_3$ ($f_3$=150 $mm$) respectively. $ \rvert V\rangle$ and $\rvert H\rangle$ represent the vertical and horizontal polarizations. $(P_1)$ and $(P_2)$ represent the image plane and $(P')$ the Fourier plane. D is a dichroic mirror, $F_1$ and $F_2$ are the interferential filters.}\label{fig4}
\end{figure}
The experimental setup in near-field is similar to that described in Fig. 2a,  when the $4-f$ imaging system is replaced by a $2-f$ imaging system ($L_1$:$f_1$=150 $mm$). In this case the position of the diffuser lies in the far-field of the crystal and the EMCCD cameras image the near-field. The transverse coordinates $x'$ and $y'$ in the near-field plane are related to the transverse coordinates in spatial frequency units $\nu_x$ and $\nu_y$ in the far-field plane by $x=\lambda f_{1}\nu_x$ and $y=\lambda f_{1}\nu_y$ , where $\lambda=710\,nm$ is the signal or idler wavelength and $f_1$ the focal length of the far-field imaging lens.
As we did before, first we remove the diffuser in the experimental setup (see Fig. 4) and we measure the correlation function with a set of 100 twin images.

\begin{figure}[htbp]
	\centering
	\includegraphics[width=14cm]{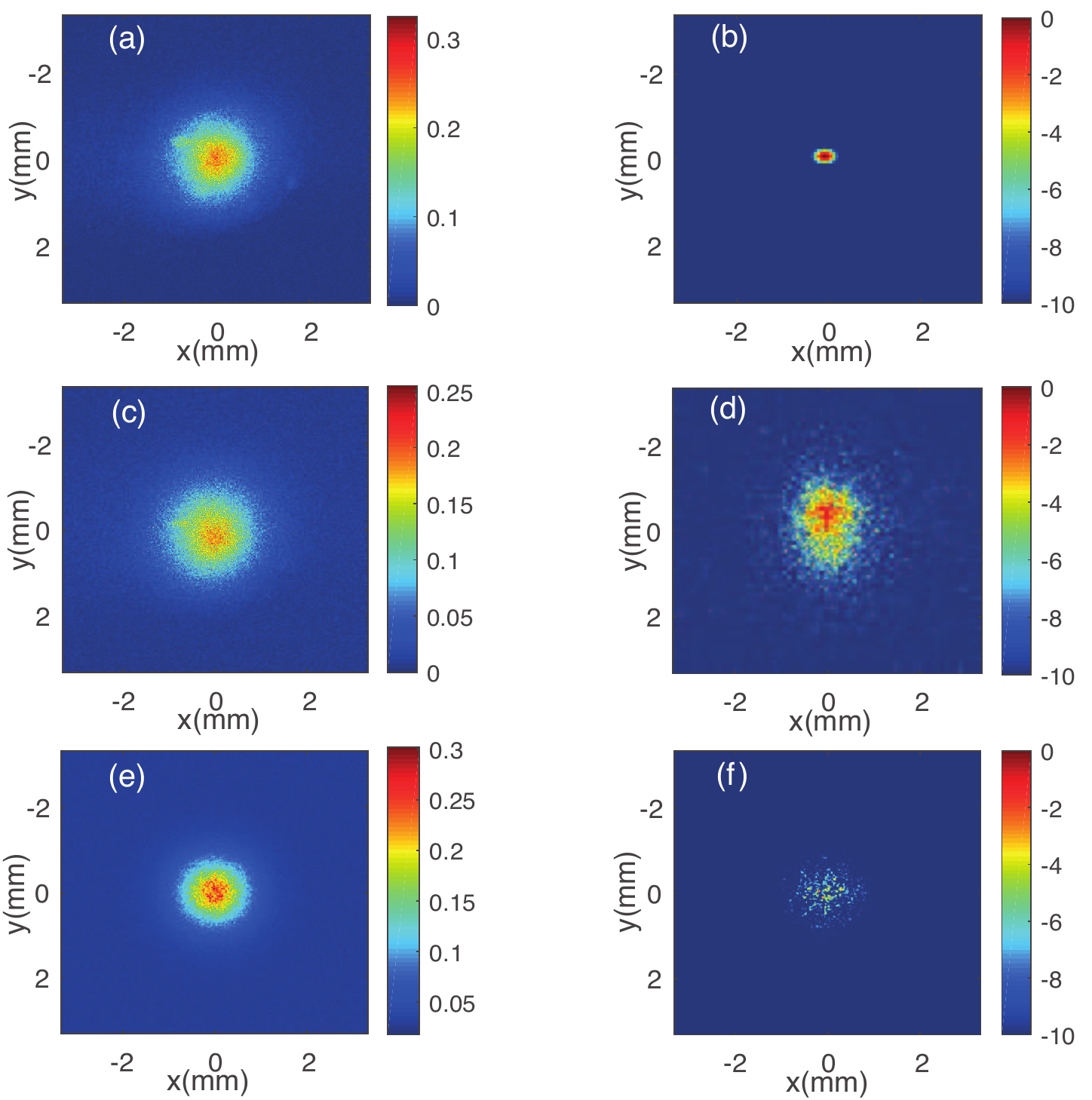}
	\caption{Without diffuser, (a) average photon number in single near-field images(signal or idler) of SPDC and (b) measured correlation function in dB between 100 twin images. With diffuser, (c) average photon number in single near-field images (signal or idler)of SPDC and (d) measured correlation function in dB over 70 000 twin images. With stochastic simulations, (e) average photon number in single near-field images (signal or idler) of SPDC and (f) correlation issued from 10 000 stochastic simulations with diffuser.}\label{fig5}
\end{figure}

Fig. 5a shows a single near-field image of the SPDC conditioned by both phase matching and the pump beam size. As written in section III.A and in agreement with \cite{peeters_observation_2010}, the absence of any speckle pattern is consistent with the incoherent character of the light formed by a single beam of the entangled light. Similarly to the far-field correlation (see Fig. 3b), near-field correlations without diffuser (see Fig. 5b) show a narrow peak. The degree of correlation deduced from this peak is equal to 0.23, slightly above that obtained in Fig. 3b. Similarly to the far-field correlations, we calculated the conditional variances in near-field using Fig. 5b. We obtained $\sigma_{x}^2=171\,\mu m^2$ along $x$ axis and $\sigma_{y}^2=72\,\mu m^2$ along $y$ axis. Using the conditional variances both in near-field and far-field, we can estimate the Schmidt number of the biphoton state in both transverse dimensions as \cite{moreau_einstein-podolsky-rosen_2014}:

\begin{equation} \label{eq10}
V_{x,y}=\frac{0,25\hbar^2}{\sigma_{\nu_{x,y}}^2\sigma_{\nu_{x,y}}^2}
\end{equation}

Using Eq.(10), we calculate the Schmidt number of $V_x=29$ and $V_y=347$ along $x$ and $y$ dimensions respectively. The dimensionality of the entanglement, or Schmidt number $V$, can be assessed as the square  root of the product of Schmidt number in each direction by the relation \cite{moreau_einstein-podolsky-rosen_2014}:

\begin{equation} \label{eq11}
V=\sqrt{V_{x}V_y}=100
\end{equation}

The Schmidt number in the two-dimensional transverse space deduced from Eq.(11) is equal to 100. This value characterizes the high-dimensionality of the spatial entanglement in our experiment, allowing coincidence imaging of quite complex objects \cite{devaux_quantum_2019}.
We now put the diffuser in the setup and detection occurs in near-field. Fig. 5c and Fig. 5e depict the single near-field images of the SPDC. Like the far-field correlation function (see Fig. 3d and Fig. 3f), the near-field correlation function (Fig. 5d and Fig. 5f) exhibits a speckle pattern. Therefore, experimental results and stochastic simulations show that the spatial structure from the magnitude of a triple correlation function in Eq. (9) could be a speckle pattern. The degree of correlation between the entangled photon pairs using a diffuser is estimated to 0.16. This value of degree of correlation is close to that obtained in far-field correlations (see Fig.3d). Moreover, experimental results and stochastic simulations show that the speckle pattern size in near-field (Fig. 5d and Fig. 5f) is related to the pump beam profile (Fig. 5c and Fig. 5e) in agreement with the fact that near-field correlations depend of the pump profile and the phase matching function. The whole speckle pattern size $\delta x$ along $x$ axis is estimated to 1.4 $mm$ (see Fig. 5d). This value of the speckle pattern size in near- field is close to pump beam profile (1.6 $mm$ FWHM beam diameter).

\section{conclusion}
We have experimentally studied spatial correlations, imaged by two EMCCD cameras, of a biphoton state transmitted by a scattering medium lying either in the far-field or in the near-field of entangled photons source. As expected, our studies have shown the absence of one photon speckle in the quantum light transmitted through a scattering medium. For the entangled photon pairs, near-field and far-field spatial correlations have been evidenced on the whole set of photons. In both configurations, the correlation function of the entangled photon pairs exhibits a speckle pattern. Our results are in perfect agreement with \cite{peeters_observation_2010} where the correlations were obtained by scanning two single detectors. No surprisingly, the estimation of the degree of correlation without diffuser and with diffuser has shown that a part of the correlations are lost in presence of a scattering medium. Since scattering of the entangled photons scrambles the quantum correlations, new technologies using spatial light modulator (SLM) are available to control quantum state of the light in random media \cite{Lib:19,defienne_adaptive_2018}.

\section*{Funding}
This work was partly supported by the French "Investissements d'Avenir" program, project ISITE-BFC (contract ANR-15-IDEX-03) and the RENATECH network and its FEMTO-ST MIMENTO technological facility.

\end{document}